# LARGE MODE AREA LEAKY OPTICAL FIBRE FABRICATED BY MCVD

B. Dussardier[1*], S. Trzésien[1], M. Ude[1], V. Rastogi[2], A. Kumar[2], and G. Monnom[1]

[1] Lab. de Physique de la Matière Condensée, Univ. de Nice Sophia Antipolis, CNRS, UMR 6622, F-06108 Nice Cedex, France

[2] Department of Physics, Indian Institute of Technology Roorkee, India

*email: bernard.dussardier@unice.fr

**Abstract:** A large mode area single-mode optical fibre based on leaky mode filtering was prepared by MCVD. The cladding structure discriminates the fundamental mode from the higher order ones. A preliminary version has 25-μm core diameter and 0.11 numerical aperture. A Gaussian-like mode with 22-μm MFD is observed after 3-m propagation, in agreement with modeling.

## 1. INTRODUCTION

Single-mode Large Mode Area (LMA) fibres are of great interest for high power transportation due to their high threshold of nonlinearities, and for the high energy storage capacity in applications such as fibre lasers and power-amplifiers. They also have potentials in carrying of high power beams over long distances and/or with narrow spectral bandwidth. In LMA fibres light guidance via total internal reflection or photonic bandgap is usually reported. Fabrication techniques include Modified Chemical Vapor Deposition (MCVD) [1,2], external direct nanoparticle depositon [3] and finally air-silica microstructuration [4]. Few reports concern so-called all-solid fibres, made only of silica due to stringent specifications on the control of the core refractive index [2].

To obtain rigourous single mode operation, irrespective of the fabrication technique, the most difficult specification to meet is the stringent control of the refractive index difference between core and cladding over a large diameter and/or of the cladding microstructure. These stringent specifications impose complex fabrication steps in order to achieve very low core-cladding index difference, and hence a numerical aperture (NA) as low as 0.04 to obtain a really single mode 20 μm-diameter core, for example.

Recently, new single mode LMA structures were proposed, which were composed of a multimode core and a structured cladding acting as a modal filter and guiding through total internal reflection of leaky modes. The cladding structuring may be azimuthal [5,6] or radial [7]. The most relevant parameter allowing effective single mode operation is the strong discrimination between the propagation loss of the fundamental mode and any higher order mode (HOM). The discrimination must be effective within a length of fibre compatible with the expected application. For example, in applications of kilometer-long beam transportations, a 1-dB/km loss on the leaky fundamental mode may be acceptable. Instead, for amplifiers and laser applications, a loss less than 0.1 dB/m is acceptable.

An important advantage of this type of fibre is that the imposed fabrication constraints on the size and shape of the core index of refraction are sensibly relaxed. A numerical study has shown that an effective mode area as large as 1000 μm$^2$ could be achieved in such a fibres using a standard and simple fabrication technique such as MCVD [8].

In this paper, we report on the preparation by MCVD and the characterization of an optical fibre designed using the principle detailed in [8]. To experimentally validate the proposed principle, we have designed a simplified refractive index profile (RIP) characterized by two annular, low refractive index trenches in the cladding surrounding a multimode core. We have prepared the fibre, characterized the mode filtering action of the cladding and measured the loss of the fundamental mode. We have observed numerically and experimentally that the effective single mode guidance is achieved beyond 2 meter long propagation.

## 2. FIBRE DESIGN AND FABRICATION

A preform with diameter 11.4 ± 0.2 mm was fabricated by MCVD (Fig. 1) and drawn into a 110.0 ± 0.2 μm-fibre. The fibre had a similar RIP as that measured in the preform, i.e. no noticeable distortion was observed. The core and other parts with same refractive index ($n_1$) are of pure silica whereas the low refractive index trenches are doped with fluorine (F). $n_2$ is the refractive index of the zone with highest F-content, in direct contact with the core. The relative refractive index difference $\Delta_1 \sim (n_1 - n_2)/n_1 = 0.28$ % at 1.55 μm ($n_1 = 1.444$) is relatively low. The fluorine concentration could be increased to obtain $\Delta \sim 0.5$ % (by MCVD). Plasma Chemical Vapour Deposition could be also used to increase Δ up to 2 %.

The diameter and the numerical aperture of the core, bounded by the first trench with index $n_2$, are 25.4 ± 0.3 μm and 0.11 ± 6%, respectively. Note that the numerical aperture is relatively high for a single mode LMA fibre. With such characteristics, the normalized frequency $V \sim 5.6$ at 1.55 µm, and five or six LP-modes would be guided. The second low refractive index trench ($\Delta_2 \sim 0.1\%$) is 25 µm away from the fibre axis, and has a ~3 µm width. Note that for applications requiring a high rare-earth doping in the core, this would increase the refractive index of the core above $n_1$. However, this increase would not disturb the filtering as long as the HOM effective indices stay lower than $n_1$, which could be easily achieved.

Numerical simulations of mode propagation on an idealized structure (Fig. 1, bottom) were performed using the transfer matrix method [8,9]. The calculated intensity diameter (at $1/e^2$ of maximum intensity) of the $LP_{01}$ mode is found at 26 µm (corresponding to an effective area equal to 530 µm$^2$). The calculated propagation loss for the $LP_{01}$ and $LP_{11}$ modes are 0.35 and 8.3 dB/m, respectively. All other HOM are expected to suffer higher loss compared to the $LP_{11}$ mode. Therefore, after three meters of propagation, the power carried by the $LP_{11}$ mode would be less than –20 dB compared to that of the $LP_{01}$ mode.

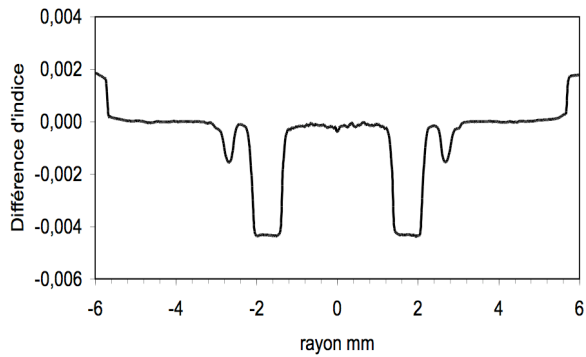

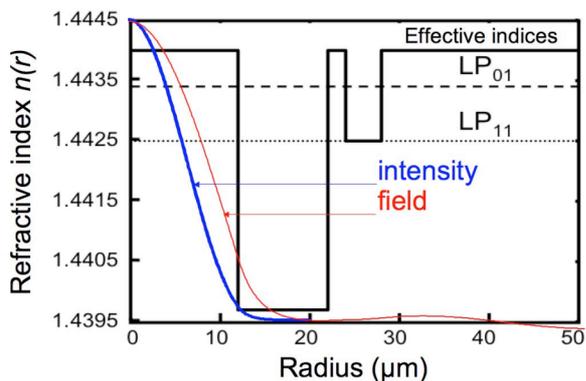

*Fig. 1 : Top : measured RIP of the preform. Bottom : modeled RIP (black), calculated fundamental $LP_{01}$ mode intensity (blue) and field (red) profiles. Horizontal dashed and dotted lines correspond to the effective index of modes $LP_{01}$ and $LP_{11}$ at 1.55 µm, respectively.*

## 3. MODAL CHARACTERIZATION

The characterization of the modal properties of the fibre at 1.55 µm was performed by exciting all possible modes with a strongly convergent beam obtained from a fibre pigtailed laser diode and a high numerical aperture lens (x 60). The output near-field was observed using another x60-microscope lens and imaged on a camera. All intensity profiles transverse dimensions have been calibrated using the same apparatus and a standard single-mode fiber (SMF, Draka, type G.652.B, mode diameter 10.1 ± 0.5 µm at 1.55 µm).

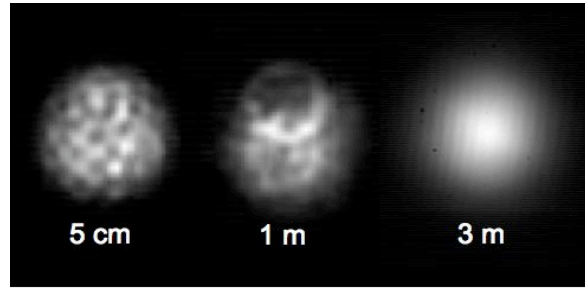

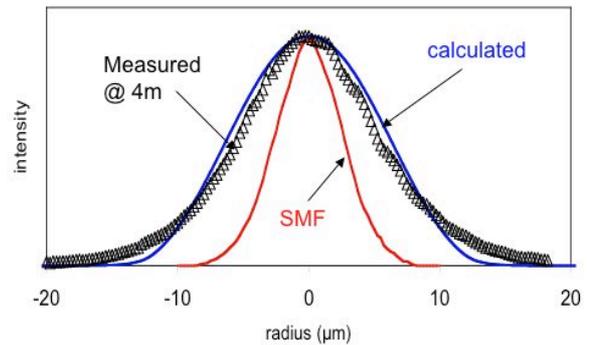

*Fig. 2 : Top: near-field intensity mode images vs length at 1.55 µm. Bottom: Intensity profiles (triangles: measured from LMA after 4 m, blue: numerical simulation, red: measured form a standard SMF).*

Images and intensity profiles were collected for various lengths of fibre, using the cut-back method. Some examples are shown on Fig. 2. For 5 cm, the multimode excitation is visible ; this image was extremely sensitive to launching conditions and fibre bending. When the fibre was longer, the profiles became smoother and their sensitivity to launching conditions diminished. When the length is higher than 2 m, the profile becomes Gaussian-like, and the launching conditions influence only the output power, not the profile. The measured $1/e^2$-mode diameter after 4 m was measured at 22 ± 2 µm (effective area ~380 µm$^2$), in relatively good agreement with numerical simulations (Fig. 2). The measured and simulated mode profiles are very similar.

## 4. LOSS MEASUREMENTS

The propagation loss of the fundamental $LP_{01}$ mode was measured using both single- and multiple cut-back methods. A fibre pig tailed tunable laser diode was connected to the input of the test fibre. A 4 cm-loop was made on the input end of the fibre to reject more strongly any HOM and therefore preferentially excite the fundamental $LP_{01}$ mode. The output power measurements were done using a power-meter.

The single cut-back technique consists in measuring the output power for two different fibre lengths (without changing the coupling conditions). Here the shortest length was larger than 3 m, and the cutback length was 1.53 m. The loss coefficient was calculated assuming exponential power decay along the fibre. The multiple cutback technique is similar, except that the fiber is cut several times (here, typically from 50 to 80 cm each time) and the output power is measured every time. The starting length was 8.78 m, and the shortest was 2.0 m, therefore more than ten measurements could be done. The loss coefficient was obtained by fitting the curve of output power versus length. The results of both series of measurements are shown on Fig. 3.

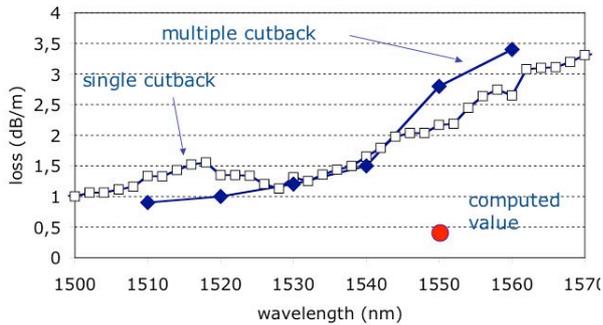

*Fig. 3 : Loss of the fundamental mode versus wavelength. Open squares: single cutback with short (and long) fibre length of 1.53 m (3 m), respectively. Filled diamonds: multiple cutback with length from 8.78 downto 2.0 m. Red disk: calculated loss at 1.55 µm on ideal RIP of Fig. 1.*

The loss of the $LP_{01}$ mode increases with the wavelength, from ~1 dB/m at 1500 nm up to 3 dB/m at 1560-1570 nm. There is a strongest loss increase from 1540 nm, onwards, showing the increase of the confinement loss of the $LP_{01}$ mode. This was expected because as the wavelength increases, the effective index of the $LP_{01}$ mode is lowering and getting closer to the index of the bottom of the second trench ($n$ = 1.4425, see bottom of Fig. 1). Therefore above a certain wavelength, even the fundamental mode suffers strong confinement loss.

The differences between both curves (typically 0.5 dB/m) are assigned to the strong sensitivity of this particular fibre to macro- and micro-bending. The calculated loss at 1.55 µm (red disk in Fig. *3*) is 2.5 dB/m lower than the measured multi-cutback one at the same wavelength. This discrepancy is attributed to several possible causes: (1) the mode confinement in the fibre is not as good as in the ideal RIP, due to imperfections in the real RIP, and (2) micro-bending losses may be high due to the fact that the RIP design is extremely simple, not yet optimized against this kind of loss. It is believed that a more complex structure in the cladding, involving a larger number of low-index trenches, would lower the confinement loss of the $LP_{01}$ mode, as already shown by numerical simulations.

## 5. DISCUSSION AND CONCLUSION

The single-mode effective operation of this fibre, having a very simple RIP structure, is in good agreement with numerical simulations. It is experimentally shown that a large and high NA core can support only the $LP_{01}$ mode after a couple of meters of propagation. A mode effective area of 380 µm$^2$ was obtained. The higher than expected confinement loss is attributed to differences between the idealized and real RIP.

A more complex version of the RIP, comprising up to four trenches with low-refractive index would allow a mode effective area up to 1000 µm$^2$ together with 40 dB-mode discrimination between $LP_{01}$ and $LP_{11}$ modes [8].

Advantages of this kind of RIP structure are many in laser and amplifiers applications : low bending loss are achievable thanks to the relatively high numerical aperture of the central core; preforms and fibre are easy to prepare because standard fabrication techniques can be implemented, such as MCVD or OVD; it is possible to dope the core with rare-earth ions for amplification; and the waveguide geometry could be strictly cylindrical for applications where low birefringence is required.

Cladding pumping for high power amplification should not be altered by the structured cladding, because the latter would be transparent to short wavelength pump modes. Therefore this class of single-mode, solid-core, large mode area optical fibres have interesting applications perspectives.

## ACKNOWLEDGEMENT

This study was partially supported by an Indo-French "*Projet de Recherche en Réseau*" (P2R) « R&D on Specialty Optical Fibers and Fiber-based Components for Optical Communications » funded